# Strain tunable pudding-mold-type band structure and thermoelectric properties of SnP$_3$ monolayer

Shasha Wei, Cong Wang and Guoying Gao[*]

School of Physics, Huazhong University of Science and Technology, Wuhan 430074, China

[*]E-mail: guoying_gao@mail.hust.edu.cn

**Abstract**

Recent studies indicated the interesting metal-to-semiconductor transition when layered bulk GeP$_3$ and SnP$_3$ are restricted to the monolayer or bilayer, and SnP$_3$ monolayer has been predicted to possess high carrier mobility and promising thermoelectric performance. Here, we investigate the biaxial strain effect on the electronic and thermoelectric properties of SnP$_3$ monolayer. Our first-principles calculations combined with Boltzmann transport theory indicate that SnP$_3$ monolayer has the "pudding-mold-type" valence band structure, giving rise to a large $p$-type Seebeck coefficient and a high $p$-type power factor. The compressive biaxial strain can decrease the energy gap and result in the metallicity. In contrast, the tensile biaxial strain increases the energy gap, and increases the $n$-type Seebeck coefficient and decreases the $n$-type electrical conductivity. Although the lattice thermal conductivity becomes larger at a tensile biaxial strain due to the increased maximum frequency of the acoustic phonon modes and the increased phonon group velocity, it is still low, only e.g. 3.1 Wm$^{-1}$ K$^{-1}$ at room temperature with the 6% tensile biaxial strain. Therefore, SnP$_3$ monolayer is a good thermoelectric material with low lattice thermal conductivity even at the 6% tensile strain, and the tensile strain is beneficial to the increase of the $n$-type Seebeck coefficient.

# 1. Introduction

Thermoelectric materials have attracted increasing interest due to their applications in energy converters and thermoelectric refrigeration. The efficiency of thermoelectric conversion is mainly governed by the dimensionless figure of merit $ZT = S^2\sigma T/\kappa$ [1, 2], where $S$ is the Seebeck coefficient, $\sigma$ is the electrical conductivity, $T$ is the absolute temperature, and $\kappa$ is the thermal conductivity including electronic ($\kappa_e$) and lattice ($\kappa_l$) contributions. The higher the $ZT$ value, the better the thermoelectric performance. However, the $ZT$ value cannot be arbitrarily increased, because these thermoelectric coefficients are connected each other, e.g. the electrical conductivity and electrical thermal conductivity are related by the Wiedemann-Franz law: $\kappa_e=L\sigma T$, where $L=2.45\times10^{-8}$ W$\Omega$K$^{-2}$ is the Lorentz number for free electrons [1, 3], and the Seebeck coefficient is inversely proportional to the carrier concentration, while the electrical conductivity is proportional to the carrier concentration. So, a semiconductor with a moderate energy gap will have a better thermoelectric performance. Usually, band engineering, doping, pressure and superlattice structure were used to improve the thermoelectric performance [4-7].

In recent years, low-dimensional especially 2D atom-layered semiconductors have been extensively studied due to their versatile electronic, optoelectronic and electrochemical properties superior to their bulks [8-10]. 2D thermoelectric properties have also become more and more popular, because the 2D quantum confinement will give rise to the local electronic density of states and increase the phonon scattering compared to the bulk, and in turn the large Seebeck coefficient and the low phonon thermal conductivity will be achieved. For example, the Seebeck coefficient of monolayer $TiS_2$ becomes 40% larger than that of the bulk [11]. The power factors for few atom-layers of transition metal dichalcogenides ($MX_2$) such as $MoS_2$, $MoSe_2$, $WS_2$ and WSe$_2$ can be greatly increased compared to their bulks due to the near degeneracy of band valleys in the 2D structures [12]. Although the lattice thermal conductivities of some MX$_2$ can be reduced from the bulk to the monolayer due to the increasing phonon scattering, these values are still high due to the high phonon frequency and the large gap between the acoustic and optical modes [13], e.g. the lattice thermal conductivities of 2H-MoS$_2$ and 2H-WSe$_2$ reach about 100 and 40 Wm$^{-1}$K$^{-1}$ at room temperature, respectively [14, 15]. Remarkably, different to 2H-MX$_2$ monolayers, 1T-MX$_2$ monolayers such as ZrSe$_2$, HfSe$_2$ and SnSe$_2$ have lower phonon frequency and coupling between the acoustic and optical modes, and thus lower lattice thermal conductivities were found such as 1.2 Wm$^{-1}$K$^{-1}$ for ZrSe$_2$, 1.8 Wm$^{-1}$K$^{-1}$ for HfSe$_2$ and 3.27 Wm$^{-1}$K$^{-1}$ for SnSe$_2$ at room temperature [16-18]. Till now, 2D atom-layer materials with low lattice conductivity are still few, and it is necessary to search for novel 2D materials with high power factor and low lattice thermal conductivity.

Recently, GeP$_3$ and SnP$_3$ attracted more and more attentions [19-24], because their bulks, which have been synthesized many years ago, exhibit metallicity [25, 26], while their monolayers and bilayers behave semiconductor properties with small indirect gap and high carrier mobility similar to phosphorene [19-24]. Importantly, SnP$_3$ monolayer was found to possess a low lattice thermal conductivity of ~4.97 Wm$^{-1}$K$^{-1}$ at room temperature due to the low acoustic group velocity, strong dipole-dipole interactions and strong phonon-phonon scattering [27]. Interestingly, for SnP$_3$ monolayer [23], a transition from an indirect to a direct semiconductor can be achieved at the 4% compressive biaxial strain, and metallicity emerges when the compressive biaxial strain is larger than 6%. In contrast, the band gap is increased under the tensile biaxial strain.

Therefore, the excellent tunable electronic structure of SnP$_3$ monolayer will broaden the possible 2D electronic, optoelectronic and thermoelectric properties. Although the thermoelectric properties of SnP$_3$ monolayer have been studied [27], to the best of our knowledge, there is no report on the strain effect on the thermoelectric performance. In this article, we use the first-principles calculations and Boltzmann transport theory to investigate the electronic structure and the electron and phonon transport properties of SnP$_3$ monolayer under a strain. It is found that the "pudding-mold-type" valence band structure leads to a large *p*-type Seebeck coefficient and a high *p*-type power factor. The tensile biaxial strain can increase the *n*-type Seebeck coefficient and decrease the *n*-type electrical conductivity. Although the lattice thermal conductivity becomes larger at a tensile biaxial strain, it is still low with 3.1 Wm$^{-1}$ K$^{-1}$ at room temperature with the 6% tensile biaxial strain.

## 2. Computational methods

The present calculations include three parts. Firstly, we optimize the structure and calculate the electronic energy band for SnP$_3$ monolayer by using first-principles density functional theory (DFT) in conjunction with projector-augmented-wave (PAW) pseudopotentials, as implemented in the Vienna ab initio simulation package [28]. The electronic exchange correlation functional is used within the generalized gradient approximation (GGA) with the Perdew–Burke–Ernzerhof (PBE) [29]. We use 20×20×1 Monkhorst-Pack *k*-point grid, and the plane-wave energy cutoff is set to 500 eV. The convergence threshold for the electronic self-consistent iteration is specified as 10$^{-6}$ eV, and only if all the forces are smaller than 0.01eV/Å, all the atomic positions and the lattice parameters are fully optimized. Because of the underestimation of the band gap within the GGA-PBE, we also use the Heyd−Scuseria−Ernzerhof (HSE06) functional [30] for the band structure calculations for comparison.

Secondly, based on the obtained electronic band structure, the electronic transport coefficients, i.e. the Seebeck coefficient $S$ and the electrical conductivity with relaxation time $\sigma/\tau$, are calculated by the BoltzTraP code [31], which is based on the semi-classical Boltzmann transport theory under the constant relaxation time and rigid band approximations. Here, it is difficult to accurately obtain the electronic relaxation time $\tau$, although it can be estimated by the deformation potential theory [32]. We mainly discuss the effect of strain on the thermal transport coefficients.

Thirdly, for the phonon transport calculations, we use the density functional perturbation theory (DFPT) in combination with VASP to calculate the force constant matrices [28]. The phonon dispersion and the phonon thermal conductivity are calculated by utilizing the Phonopy package [33] and by solving the phonon Boltzmann transport equation as implemented in the ShengBTE [34], respectively. The second order harmonic and third order anharmonic interatomic force constants (IFCs) are calculated by using a 4×4×1 supercell containing 128 atoms based on the relaxed unit cell. Besides, the harmonic IFCs are obtained by using the Phonopy package [33], and the anharmonic IFCs are calculated by considering the interactions up to third-nearest neighbors. The present computational methods of electron and phonon transport coefficients have been successfully used in our previous works of bulk and low-dimensional systems [16, 17, 35].

## 3. Results and discussion

**3.1. Electronic band structure and strain effect**

Bulk $SnP_3$ has the layered structure with the trigonal space group of R-3m [25]. Monolayer $SnP_3$ has the hexagonal lattice with the space group of P-3m1, which is show in figures 1(a) and 1(b). Our optimized equilibrium lattice parameter a=b=7.15 Å is in good agreement with the previous values of 7.105, 7.16 and 7.15 Å [22-24], and it is a little smaller than the in-plane value of bulk SnP$_3$ of 7.378 Å [25]. The calculated electronic band structure at equilibrium lattice is given in figure 1(c). For comparison, both results at GGA-PBE and HSE06 levels are presented, because the HSE06 functional can be used to correct the underestimation of the energy gap caused by GGA-PBE or LDA functional. Figure 1(c) indicates that monolayer SnP$_3$ is a semiconductor with an indirect band gap of 0.43 eV (GGA-PBE functional) or 0.68 eV (HSE06 functional), which is consistent with the previous results of 0.39 and 0.43 eV at GGA-PBE and 0.67 and 0.72 eV at HSE06 [23, 22]. Interestingly, the valence bands around the Fermi level exhibit a "pudding-mold-type" band structure [36-38], because the valence bands along the M-K direction are flat, but the ones along the M-Γ and K-Γ directions are dispersive. These characteristics will lead to a large $p$-type Seebeck coefficient and a large $p$-type electrical conductivity and in turn a high power factor. Comparing the band structure and the total and atomic orbital density of states (DOS) (figure 1(d)), one can see that the "pudding-mold-type" valence bands are mainly contributed by the P-3$p$ and Sn-5$p$ orbitals, and the conduction band around the Fermi level mainly originates from the Sn-5$p$ orbitals.

We now discuss the biaxial strain effect on the electronic band structure of monolayer SnP$_3$. Note that the strain effect is considered within GGA-PBE, because the change tendency of band structure with strain is usually same to that within HSE06, and the gap variation trends with strain have been successfully predicted with the GGA-PBE level [20, 39]. To simulate the biaxial strain, we use the formula of $\varepsilon=\Delta a/a_0 \times 100\%$, where $a_0$ is the optimized equilibrium lattice constant, and $\Delta a$ is the change of the lattice constant under a biaxial strain. Figure 2 shows the calculated electronic band structure under compression biaxial strains of -3% and -6% and tensile biaxial strains at 3% to 6%. It is clear that the band gap decreases with the increasing compressive strain, and the valence band maximum moves from K to Γ, leading to the transition of indirect-to-direct energy gap. When the compressive strain reaches 6%, the conduction band around the Fermi level becomes more dispersive and touches the Fermi level, making monolayer $SnP_3$ become a metal. For the tensile strain, the band gap becomes large with increasing strain. The conduction band becomes more flat, while the valence bands along the K-Γ direction become more dispersive. All these changes with strain are in agreement with the previous results [23, 22].

**3.2. Electronic transport with tensile strain**

Figure 3 shows the calculated carrier-concentration- and temperature-dependent Seebeck coefficient $S$, electrical conductivity with relaxation time $\sigma/\tau$ and power factor with relaxation time $S^2\sigma/\tau$. One can see that the $p$-type Seebeck coefficient is much higher than that of the $n$-type. The reason is that the Seebeck coefficient is proportional to the band effective mass, and the "pudding-mold-type" valence band structure gives rise to a better flat valence band around M-K and in turn a high band effective mass. Simultaneously, the "pudding-mold-type" valence band structure has the similar dispersion to the conduction band along the M-Γ and K-Γ directions, and thus the difference of $\sigma/\tau$ between $p$-type and $n$-type is slight (figure 3(b)). Therefore, the $p$-type has much higher power factor than the $n$-type (figure 3(c)). Figure 3 also indicates that the power factor especially in the $p$-type can be greatly increased with the increasing temperature, which mainly originates from the increasing Seebeck coefficient, because the change of electrical conductivity with temperature is small.

Because monolayer $SnP_3$ becomes a metal when the compressive biaxial strain is larger than 6%, we mainly focused on the tensile biaxial strain effect on the thermoelectric properties. Figure 4 gives the changes of carrier-concentration-dependent thermoelectric coefficients at 300 K without and with 3% and 6% tensile strain. For the *p*-type doping, with the increase of tensile strain, the Seebeck coefficient decreases while the electrical conductivity increases, this is because the valence band between Γ and K becomes more dispersive (figure 3(b)), decreasing the band effective mass and increasing the carrier mobility. In contrast, the conduction band around the Fermi level becomes more flat with increasing tensile strain, increasing the band effective mass and decreasing the carrier mobility, and turn, the *n*-type Seebeck coefficient and electrical conductivity are increased and decreased, respectively. Combing the opposite changes of the Seebeck coefficient and the electrical conductivity, under a tensile strain, the *p*-type power factor is decreased, and the *n*-type power factor is increased only in the small range of carrier concentration around $7.3 \times 10^{19} \sim 3.4 \times 10^{20}$ cm$^{-3}$ (figure 4(c)). Similar changes of these electronic transport coefficients at 700 K with the tensile strain have also been found for SnP$_3$ monolayer (figure 5).

**3.3. Phonon transport with tensile strain**

We present in figure 6(a) the calculated phonon spectra without and with 6% tensile strain for monolayer SnP$_3$. There are three acoustic phonon modes (with lowest frequency) and twenty-one optical phonon modes. Usually, the main contribution to the phonon thermal conductivity is the acoustic phonon modes, and thus we mainly analyze the change of acoustic phonon modes with the strain. Some noteworthy differences of the phonon spectra between 0% and 6% tensile strains can be found. Figure 6(a) indicates that the maximum frequency of the acoustic phonon modes at 6% strain is about 1.32 THz, which is a little higher than that of 1.27 THz at 0% strain. Both values are much lower than those of 7.5 THz for MoS$_2$ monolayer and 4.8 THz for MoSe$_2$ monolayer which have higher lattice thermal conductivities of 100 and 40 $Wm^{-1}K^{-1}$ at room temperature, respectively [14, 15], and are also lower than that of 3.45 THz for $SnSe_2$ monolayer which is a good 2D material with the low lattice thermal conductivity of 3.27 $Wm^{-1}K^{-1}$ at room temperature [17, 40]. Such low frequency of the acoustic phonon modes means the phonon group velocity and the lattice thermal conductivity will be low for monolayer $SnP_3$. In addition, it is noted that the acoustic phonon modes are coupled with the optical phonon modes, which is beneficial to the increase of the phonon scattering and the decrease of the phonon thermal conductivity.

Figure 6(a) also shows that, under the strain, the acoustic phonon band with lowest frequency becomes more dispersive, indicating the increase of the phonon group velocity. Indeed, our calculated group velocities of the three acoustic phonon modes (two transverses acoustic (TA) modes and one longitudinal acoustic (LA) mode) along the Γ-M direction without strain are about 1.063, 1.342 and 1.852 km/s, respectively, and about 0.199, 0.143 and 0.601 km/s along the M-K direction. At the 6% tensile strain, the group velocities of TA and LA modes along the Γ-M are about 1.189, 1.240 and 1.846 km/s, and about 0.220, 0.172 and 0.787 km/s along the M-K direction. So, the average group velocities of the acoustic phonon modes are increased at the tensile strain, and in turn, the lattice thermal conductivity will be increased.

We further present in figure 6(b) the calculated phonon thermal conductivity with the change of temperature from 300 K to 800 K for monolayer $SnP_3$ without and with 6% tensile strain. It can be seen that the phonon thermal conductivity is low due to the low frequency of the acoustic phonon modes and the coupling between the acoustic phonon modes and the optical phonon modes. At the tensile strain, the lattice thermal conductivity is increased, as discussed above, this is due to the increase of the maximum frequency of the acoustic phonon modes and the increase of the phonon group velocity. Although the tensile strain increases the lattice thermal conductivity, but the value is still low, e.g. 3.10 $Wm^{-1} K^{-1}$ at room temperature with 6% tensile strain, which is much lower than those of 100 $Wm^{-1}K^{-1}$ for $MoS_2$ monolayer [14] and 40 $Wm^{-1}K^{-1}$ for $WSe_2$ monolayer [15], and is comparable with that of 3.27 $Wm^{-1}K^{-1}$ for $SnSe_2$ monolayer [17]. Therefore, monolayer $SnP_3$ is a good thermoelectric material with a low lattice thermal conductivity without and with tensile strain.

## 4. Conclusion

Using the first-principles calculations and the Boltzmann transport theory, we have explored the biaxial strain effect on the electronic and thermoelectric properties of $SnP_3$ monolayer, a recently discovered 2D semiconductor with high carrier mobility. Our results indicate that $SnP_3$ monolayer has the "pudding-mold-type" valence band structure with the energy gap of 0.68 eV, which results in a large *p*-type Seebeck coefficient and a high *p*-type power factor. At a compressive biaxial strain, the energy gap is decreased, and the metallicity emerges when the compressive strain is larger than 6%. In contrast, the energy gap is increased at a tensile biaxial strain, and the *n*-type Seebeck coefficient and *p*-type electrical conductivity can be improved. The low frequency of the acoustic phonon modes and the coupling between the acoustic phonon modes and the optical phonon modes make the lattice thermal conductivity low. At the tensile strain, both the maximum frequency of the acoustic phonon modes and the phonon group velocity are increased, leading to the increase of the lattice thermal conductivity. However, the lattice thermal conductivity is still low, e.g. 3.1 $Wm^{-1} K^{-1}$ at room temperature with the 6% tensile biaxial strain. Therefore, $SnP_3$ monolayer is a good thermoelectric material with low lattice thermal conductivity even at the 6% tensile biaxial strain.

## Acknowledgments

This work was supported by the National Natural Science Foundation of China under Grant No. 11474113.

# Figures

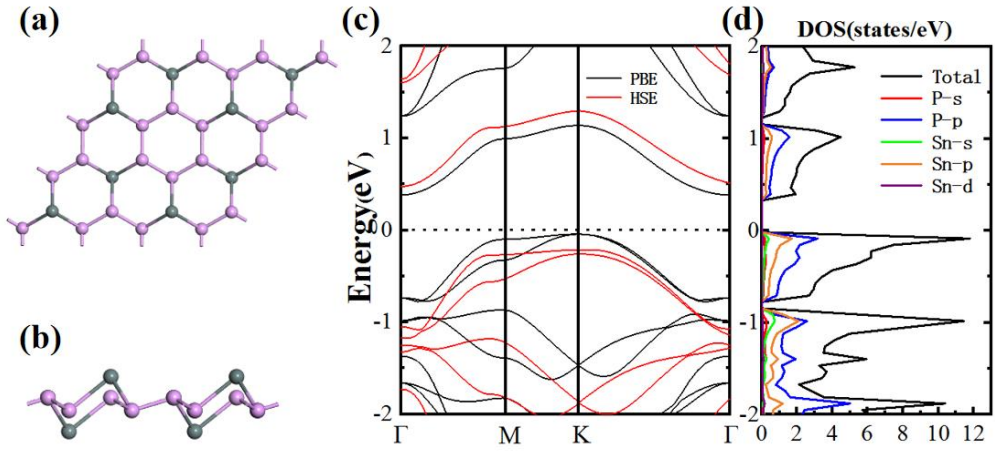

**Figure 1.** The geometry structure (top view (a) and side view (b)), energy band (c) and total and partial density of states (d) for monolayer $SnP_3$. The gray and amaranth balls represent the Sn and P atoms, respectively.

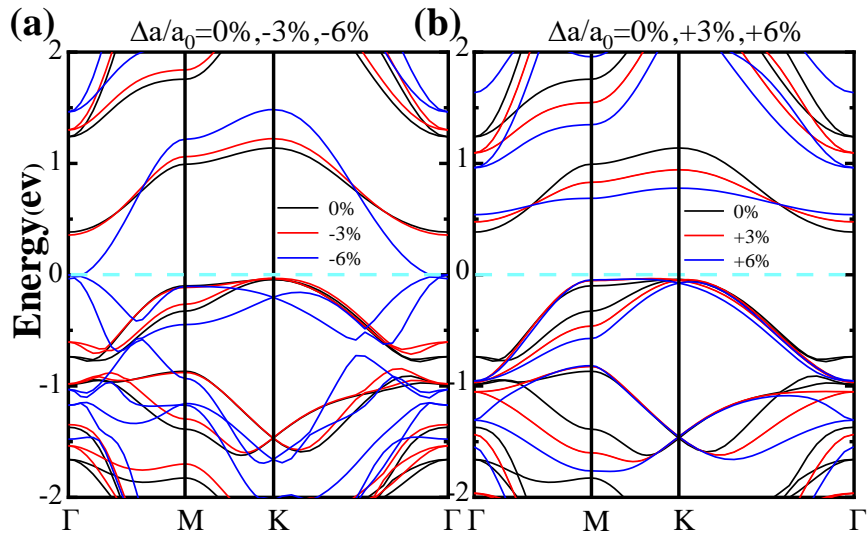

**Figure 2.** The calculated band structures of monolayer $SnP_3$ at the PBE level as a function of compressive (a) and tensile (b) strains, respectively.

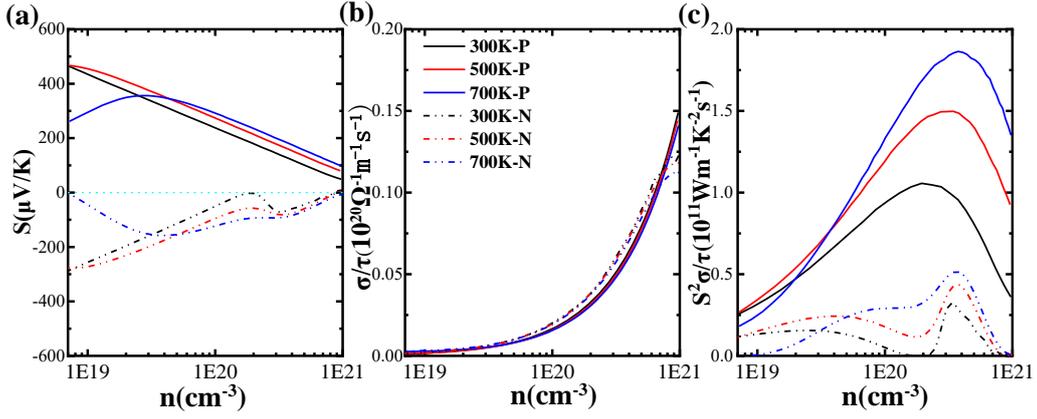

**Figure 3.** The calculated Seebeck coefficient S (a), electrical conductivity with relaxation time σ/τ (b), and power factor with relaxation time $S^2σ/τ$ (c) for both *p*-type and *n*-type monolayer $SnP_3$ as a function of carrier concentration at the temperatures of 300 K, 500 K and 700K.

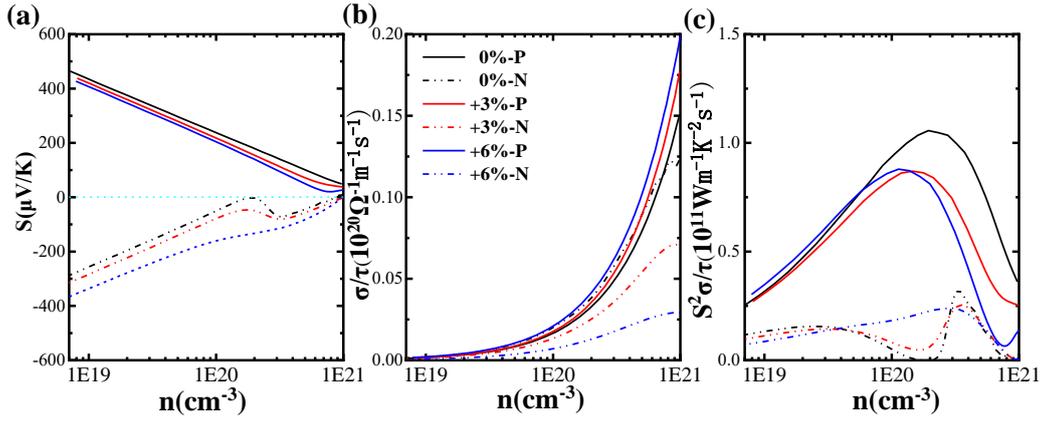

**Figure 4.** The calculated Seebeck coefficient S (a), electrical conductivity with relaxation time σ/τ (b), and power factor with relaxation time $S^2σ/τ$ (c) at 300 K for both *p*-type and *n*-type monolayer $SnP_3$ as a function of carrier concentration at the tensile strains of 0%, 3% and 6%.

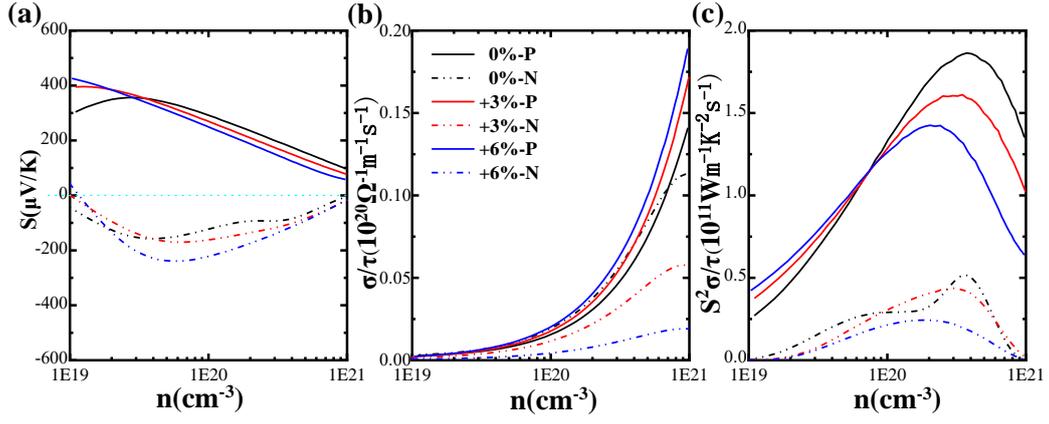

**Figure 5.** Similar to figure 4 at the temperature of 700 K.

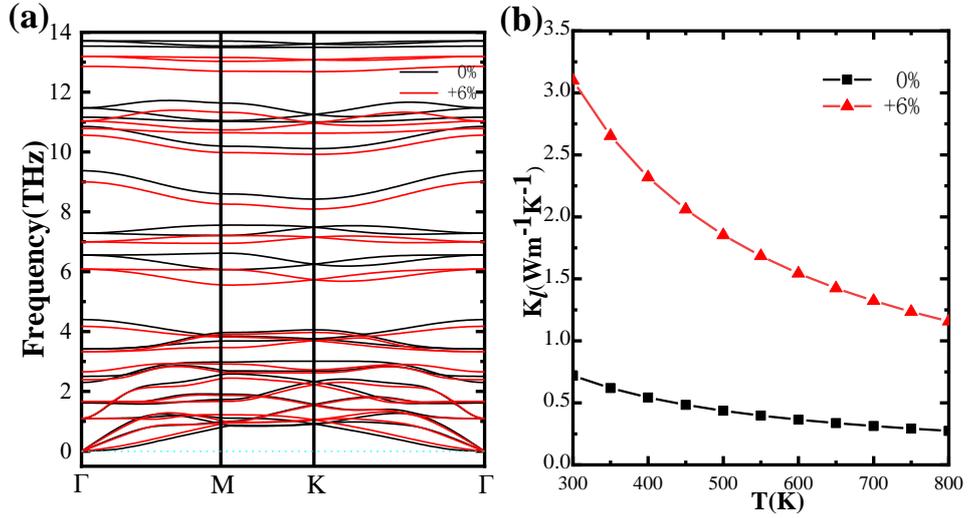

**Figure 6.** The calculated phonon spectrum (a) and lattice thermal conductivity with temperature (b) for SnP$_3$ monolayer at the 0% and 6% tensile strains.